\documentclass[prl,aps,epsf,twocolumn,showpacs]{revtex4-1}
\usepackage[pdftex]{graphicx}
\usepackage{dcolumn}
\usepackage{bm}
\usepackage{epsfig}
\usepackage{latexsym}
\usepackage{amsmath}
\usepackage{color}
\usepackage{array}

\newcommand{\up}{\uparrow}
\newcommand{\down}{\downarrow}

\renewcommand{\(}{\left(}
\renewcommand{\)}{\right)}
\renewcommand{\[}{\left[}
\renewcommand{\]}{\right]}
\renewcommand{\v}[1]{\mathbf{#1}} 

\newcommand{\br}{{\bf r}}

\begin{document}

\title{Anomalous Supercurrent from Majorana States in Topological Insulator Josephson Junctions}

\author{Andrew C. Potter}
\author{Liang Fu}
\affiliation{Department of Physics, Massachusetts Institute of Technology, Cambridge, MA 02139, USA}

\begin{abstract}
We propose a Josephson junction setup based on a topological insulator (TI) thin film to detect Majorana states, which exploits the unique helical and extended nature of the TI surface state.  When the magnetic flux through the junction is close to an integer number of flux quanta, Majorana states, present on both surfaces of the film, give rise to a  narrow peak-dip structure in the current-phase relation by hybridizing at the edge of the junction.
 Moreover, the maximal Majorana-state contribution to Josephson current takes a (nearly) universal value, approximately equal to the supercurrent capacity of a single quantum-channel. 
These features provide a characteristic signature of Majorana states based entirely on supercurrent. 
\end{abstract}

\maketitle

Majorana bound-states in superconductors are localized quasi-particles that are equal weight superposition of electron and hole, which have non-Abelian braiding statistics\cite{readgreen,ivanov}.  
The presence of Majorana bound-states can produce unusual transport phenomena such as a 4$\pi$-periodic Josephson effect\cite{kitaev, yakovenko}, resonant Andreev reflection\cite{lee}, and inherently nonlocal transport\cite{fu}. 
Over the last few years, proposals 
for realizing Majorana states in various superconducting solid-state systems have sparked tremendous interest and intensive activity (see \cite{beenakker, alicea} and references therein). Recent experiments on semiconductor nanowires with proximity induced superconductivity\cite{lutchyn, oreg} observed signatures\cite{kouwenhoven, heiblum, xu, rokhinson} interpreted as evidence of Majorana states. 

Among the proposed  material systems for realizing Majorana states, the superconductor (S)-topological insulator (TI) hybrid structure\cite{fukane} 
has several distinctive features. First,  as a parent phase for Majorana states, the S-TI interface is 
topologically nontrivial even at zero magnetic field. 
Consequently, the induced superconductivity on the topological surface states of a TI is immune to disorder\cite{potterlee}. 
This provides a robust route to realizing Majorana states at elevated temperatures.  
Second, by their topological nature, surface states of a TI extend throughout the entire sample boundary. 
The extended nature of such states has motivated us to propose a TI-based setup for  detecting Majorana states.

\begin{figure}[ttt]
\begin{center}
\includegraphics[width = 3.5in]{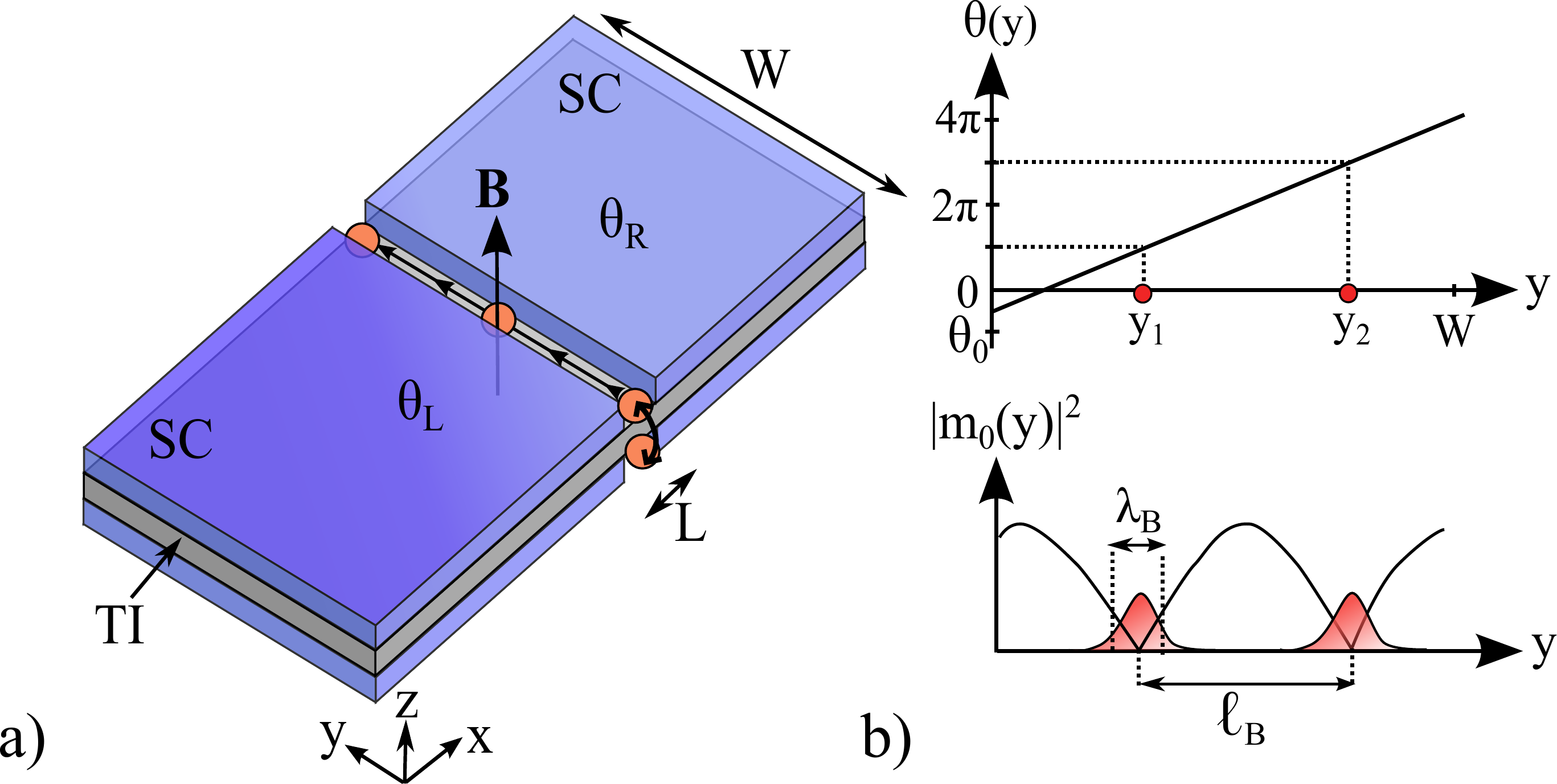}
\end{center}
\vspace{-.2in}
\caption{Panel a) shows a depiction of the conceptually simplest device geometry considered in the main text (simpler to fabricate devices such as those measured in Ref.~\cite{Sacepe, samarth, brinkman, williams, lu, Mason} are also discussed).  The top and bottom surfaces of the topological insulator (grey layer labeled TI) are in contact with superconductors (blue layers).  As the global-phase offset between the left and right superconductors ($\theta_0=\theta_R-\theta_L$) is adjusted, Majorana modes (shown as red-circles) bound to Josephson vortices are created at one end of the junction, move along the junction, and fuse on the opposite side of the junction.  Panel b) shows the local phase difference $\theta_y$ for  along the junction $3\Phi_0>\Phi_B>2\Phi_0$ and fixed $\theta_0$, and the corresponding mass term of Eq.~\ref{eq:heff}. Wherever $\theta_y = \pi\text{ mod }2\pi$, there is a local gap closing that binds a Majorana state.}
\vspace{-.2in}
\label{fig:Device}
\end{figure}

In this paper, we study anomalous Josephson current signatures of Majorana states in a superconductor-topological insulator thin film-superconductor junction under an applied magnetic field. 
Importantly, the superconductors and the TI  film are arranged in such a way that induced superconductivity 
exists on both top and bottom surfaces of the TI. 
The conceptually simplest example of such a setup is shown in Fig.~\ref{fig:Device}a, in which the superconductors wrap around the TI in the entire $y$ and $z$ direction and the supercurrent flows in the $x$ direction.  This geometry will provide a useful starting point for theoretical analysis.  However,  our results will also apply to the simpler to fabricate geometry where the superconductors are deposited only on the top surface, provided that the superconductivity is transmitted to the bottom surface through either bulk states (which are present in all experimentally realized TI materials), or side surfaces.
Supercurrent through such SC-TI-SC junctions has been recently observed\cite{Sacepe, samarth, brinkman, williams, lu, Mason}.

Our main findings and the basic physics behind them can be stated in simple terms. Applying a magnetic field to  the S-TI-S junction induces a one-dimensional array of Josephson vortices, due to the winding of the phase difference between the two superconductors, $\theta_y\propto B y$, in the $y$  direction parallel to the junction.  Each vortex traps a  localized Majorana state on the top surface of the TI, and one on the bottom surface. The two are aligned vertically (see Fig.~\ref{fig:Device}). 
The magnetic flux through the junction determines the number of Majorana states and their separation. 
The global phase offset $\theta_{y=0}$ defined at a reference point $y=0$ is an independent variable, which can be controlled in a SQUID geometry.   
For a given magnetic field, increasing $\theta_0$  shifts the positions of all Majorana states along the junction, towards one edge of the TI sample. When the first pair of Majorana states, one from the top surface and the other from the bottom, approach the edge, they become hybridized due to the wavefunction overlap on the {\it side} surface, and eventually annihilate each other.  As $\theta_0$ increases further, this fusion event repeats for the next pair of Majorana states and so on. Meanwhile, an opposite event occurs at the other edge, where pairs of Majorana states are created. Decreasing the phase difference or reversing the direction of the magnetic field will reverse the direction of motion of Majorana states and interchange the role of the two edges. 
 
Each creation or fusion event causes an energy splitting of the pair of  Majorana states involved, and for this reason, contributes to the DC Josephson supercurrent an amount that is {\it independent} of the width of the junction. 
When the magnetic flux through the junction is (close to) an integer multiple of flux quanta, 
the normal contribution to the supercurrent oscillates with position $y$ along the junction and (nearly) cancels to zero, 
while the Josephson supercurrent $I_M$ from splitting Majorana states dominates.  Advancing $\theta_0$ by $2\pi$ shifts each Majorana state to the next one. During this cycle, the first pair of Majorana states in the array is fused, and a new pair is created and takes the place of the last one. This leads to a narrow peak in Josephson current $I_M$ as a function of phase difference within each $2\pi$ period (see Fig.1d). 

\vspace{6pt}\noindent\textbf{\textit{Model - }}
We now derive the Majorana states and the Josephson current-phase relation  for a {\it short} and {\it wide} S-TI-S junction, under the condition $L < \xi < W$ where $L$ is the length of the junction, $W$ is the width, and $\xi$ is the coherence length of topological surface states with proximity-induced superconducting gap.  
Moreover, for simplicity we consider the case that 
the Josephson current is unable to screen the external magnetic field. 


The Hamiltonian for the top surface of the S-TI-S junction is 
\begin{align}
H =& \int d^2\br  \psi^\dagger(\br) \left[ v \hat{z}\cdot(\boldsymbol{\pi}\times\v{s}) - \mu(\br)  \right] \psi(\br)  
\nonumber \\
&+ \[ \Delta(\br) \psi^\dagger_\up(\br) \psi^\dagger_\down(\br) +\text{h.c.}\]
\label{h}
\end{align} 
Here, $\pi_{j} = -i \partial_{j} - e A_{j}(\br)$ ($j=x,y$), and $\psi(\br)=(\psi_\uparrow(\br), \psi_\downarrow(\br))$ describes the TI surface states, which have a Dirac dispersion  
with Fermi velocity $v$. Due to doping from the superconductors, the chemical potential 
in the superconducting region $\mu(\br)=\mu'$ at $|x|>L/2$ is different from the junction area $\mu(\br)=\mu$ at $|x|<L/2$. 
 $A_j(\br)$ is  the vector potential associated with the magnetic field $B$. 
It is convenient to work in the Landau gauge $A_x=0$. $A_y(\br)$ is then 
given by 
$A_y(\br) = -B x$ for $|x| < L/2$, $-B L/2$  for  $x > L/2$ and 
$B L/2$ for $x < -L/2$. 
In this gauge, the superconducting gap $\Delta(\br)$ is given by
 \begin{align}
 \Delta(\br) &= \left\{\begin{array}{lc}\Delta e^{ i \theta_y/2} &   x > L/2 \\
 \Delta e^{ -i \theta_y/2}& x< -L/2 \\
 0& |x|<L/2
\end{array}\right.
 \label{delta} 
\end{align}
The phase winding $\theta_y=\pi y /l_B$ ensures zero supercurrent along the $y$ direction in the superconducting region. Here, the magnetic length is defined as: $l_B = W\frac{\Phi_0}{\Phi_B}$, where $\Phi_B$ is the magnetic flux through the junction\cite{Footnote:FluxConcentration}).

Since  $\Delta(\br)$ varies slowly with the position $y$, 
we use semiclassical method to first solve the Hamiltonian  
without the kinetic energy term $- i v \partial_y s_x$ at an arbitrary $y$. 
This one-dimensional problem was previously solved for TI Josephson junction\cite{fukane} 
and for a related problem in graphene\cite{titov}. 
We quote their results below. 
When the junction length $L$ is shorter than the coherence length $\xi \equiv v/\Delta$, there is a single pair of subgap Andreev bound states 
at energy $\pm E(y)$, where $E(y)=\Delta \cos (\theta_y/2)$  depends on the local phase difference and oscillates with the position $y$ (see Fig.~\ref{fig:Device}b). 
Importantly, the two branches of Andreev bound states cross each other 
at $E(y)=0$, when the local phase difference $\theta_y$ is an odd multiple of $\pi$. 
Such a level crossing corresponds to a local change in the fermion parity over a cycle, which 
is robust against perturbations\cite{fukane2}. 

\vspace{6pt}\noindent\textbf{\textit{Majorana Bound-States - }}We now take the kinetic energy along the $y$ direction into account, to obtain the low-energy spectrum of the junction near $E=0$.    
For this purpose, it is instructive to first consider the region near a level crossing point $y=y_1$. This region can be described\cite{fukane} by the following effective Hamiltonian 
in terms of two branches of counter-propagating Majorana fermions $\gamma_{L,R}$:  
\begin{align} 
H_{\rm eff} &=  i v_M  (\gamma_L \partial_y \gamma_L - \gamma_R \partial_y \gamma_R)   +  i E(y) \gamma_L  \gamma_R, \label{eq:heff}
\end{align}
with $E(y) \approx  \pi \Delta (y-y_1) / l_B$, and 
where  the velocity $v_M$ arises from the kinetic energy along $y$ direction.  
The bound state energy $E(y)$ corresponds to the hybridization between the two Majorana states, and vanishes linearly at $y=y_1$.   

The sign reversal of $E(y)$ as a function of $y$ 
gives rise to a zero-energy Majorana bound state that is spatially localized in the junction at the position $y_0$. 
The corresponding Majorana operator $\gamma_1$ is given by:  
\begin{align}
\gamma_1 =  \int dy \frac{e^{-(y-y_1)^2/2\lambda_B^2}}{\sqrt{2\pi\lambda_B^2}}\frac{1}{\sqrt{2}}\[\gamma_L(y)+\gamma_R(y)\]\label{eq:gamma}
\end{align}  
where the decay length in the $y$ direction $\lambda_B$ is: $\lambda_B = \sqrt{v_M l_B /\pi \Delta}$. This Majorana state is confined by the TI band gap in the $z$ direction, the proximity-induced superconducting gap in the $x$ direction, and the magnetic-field-induced linear potential $E(y)$  in the $y$ direction. The decay length in the $z$ direction is given by the penetration depth of topological surface states into the bulk, which is typically a few nanometers. The decay length in the $x$ direction is given by the coherence length $\xi$ (typically a few hundred nanometers).
The velocity  $v_M$ depends on intrinsic properties of the junction.  
For a ballistic junction, it was found that $v_M \simeq v (\Delta/\mu)^2$ for $\mu=\mu'$ in Ref.\cite{fukane} 
and $v_M \simeq v (\Delta/\mu) \sin ( \mu L/v)$ for $\mu \ll \mu'$ in Ref.\cite{titov}. 
Since $\Delta/\mu$ is typically of the order $10^{-3}$, in both cases the decay length of Majorana states $\lambda_B$ along the junction is much smaller than their separation. 
Therefore the wavefunction overlap of two Majorana states along the junction is vanishingly small. For this reason, we are justified to 
treat these Majorana states as essentially non-interacting, except when a pair of them are near the edge of the TI.   

Near each $\pi$-crossing the zero-energy Majorana state is accompanied by other non-zero energy Andreev-bound-states (ABS's).  Approximating the mass term by a linear potential, $E(y) = \pi\Delta(y-y_1)/\ell_B$, (valid in the vicinity of $\theta(y)\approx\pi$), these ABS's have energies $\pm E_n$, where $E_n\approx \sqrt{2\pi n v_M\Delta/l_B}$, $n=1,2,\dots,n_\text{max}$, and $n_\text{max}\approx \ell_B\Delta/v_M$.  

\vspace{6pt}\noindent\textbf{\textit{Contrast to Conventional Junctions - }} Low-energy Andreev bound-states can also occur near $\pi$ phase difference in conventional metallic Josephson junctions under ideal conditions. For a transparent superconductor-normal interface, and in the absence of spin-orbit coupling, a conventional 2D metallic junction can be thought of as two separate copies of Eq.~\ref{eq:heff} one for right moving spin-up electrons and left-moving spin-down holes, and another for the corresponding time-reversed partners.  This would lead to low-energy ABS's bound to Josephson vortices similar to those described above, except doubly degenerate.

However, in practice normal scattering at the S-N interface (for example due to the chemical potential mismatch in S and N region) 
and spin-orbit coupling will 
push ABS up in energy towards the bulk gap $\Delta$, and even completely remove them in a short Josephson junction. 
In contrast, the helical nature of the TI surface state gives rise to topologically protected ABS's, which is also robust against disorder. 
Therefore, in the short-junction limit, the Josephson signatures described below are particular to the special properties of the S-TI-S junction.

\vspace{6pt}\noindent{\bf\textit{Josephson Current - }}
Having discussed the structure of low-energy Andreev bound-states in the junction, we now analyze their effect on Josephson current.  When the number of magnetic flux quanta, $\Phi_B/\Phi_0$, is not close to a non-zero integer then the Josephson current is carried predominately by conventional states (extended states with $E>\Delta$, and non-zero energy ABS's) and shows a nearly sinusoidal current phase dependence.  Near integer values of flux quanta, $\Phi_B/\Phi_0 = \pm 1,\pm 2, \text{etc}\dots$, however, the conventional contribution becomes vanishingly small.

In this regime, the Majorana bound-state contribution to Josephson current dominates.  For simplicity, we will first 
describe the situation for $\Phi_B=\Phi_0$ in detail.  The case of $\Phi_B=N\Phi_0$ is similar. When $\Phi_B=\Phi_0$, and $\theta_0\neq \pi$, there is exactly one Josephson vortex piercing the junction, which binds Majorana modes at $y_0 = \frac{W}{2}(1-\frac{\theta_0}{\pi})$ on both the top and bottom surfaces of the TI.  These Majorana modes hybridize by tunneling into each other around the perimeter of the junction.  The resulting energy splitting is exponentially suppressed as $e^{-2\min\[y_0,(W-y_0)\]/\lambda_B}$, and is negligibly small when the position of Majorana modes $y_0$ is more than a few $\lambda_B$ away from the edges of the junction.  In this regime, the splitting is insensitive to $y_0$ and $\theta_0$, and the Majorana modes do not contribute to the Josephson current.  

In contrast, when $\theta_0\approx \pi$, $y_0\approx 0$ and the Majorana states are strongly coupled near the edge and split away from zero-energy.  If the junction height $h\lesssim \lambda_B$, then we may ignore then finite thickness of the TI film, and the Majorana states are split by energy $\delta E_M \approx \sqrt{\frac{v_M\Delta}{\ell_B}}$.  
As $\theta_0$ approaches $\pi$ from below, the Majorana states move towards the junction edge at $y=0$ and begin to fuse and split when $\theta_0\approx \pi\(1-\frac{\lambda_B}{W}\)$.  Increasing $\theta_0$ beyond $\pi$ causes a different set of Majorana states to emerge from near $y=W$, and move to decreasing $y$, reversing the process that occurred near $y=0$ for $\theta_0>\pi$.  The hybridization of Majorana states at the two edges gives rise to local supercurrents in {\it opposite} directions. 
At $\theta_0=\pi$, the splitting of Majorana states is large, but supercurrents from two edges cancel.   
Slight deviation from $\theta_0=\pi$ tips the balance by increasing the magnitude of supercurrent at one edge and suppressing the other, 
thereby generating a nonzero total supercurrent.  
The sensitivity of the Majorana splitting energy to the phase difference implies that the Majorana states contribute a peak in the Josephson current near $\phi_0\approx \pi\(1-\frac{\lambda_B}{W}\)$ followed by a dip in the Josephson current near $\phi_0 \approx \pi\(1+\frac{\lambda_B}{W}\)$.

By these considerations, we find that the maximal supercurrent carried by the Majorana state is $I_M\approx \frac{2\pi}{\Phi_0}\frac{\delta E_M}{\delta\theta_0}\approx \Delta/\Phi_0$, which for Al SC layers is $I_M\approx 10$nA.   We note that this result is largely independent of details.  In particular it is completely independent of $v_M$, $\mu$, $W$, and $L$ (though it can depend slightly on finer details such as the degree of asymmetry between the top and bottom surfaces, or between the two edges). 
 The Majorana current $I_M\approx \Delta/\Phi_0$ roughly corresponds to the maximal amount of Josephson current carried by a single quantum channel\cite{BeenakkerSQPC}.  By comparison, for junctions of a few $\mu$m in width, such as those measured in Ref.~\onlinecite{williams}, there are roughly $ k_FW\approx 10^2 -10^3$ quantum channels in the entire junction.  Consequently, the Majorana contribution to Josephson current, which dominates near an integer flux quanta, is $10^{-2} - 10^{-3}$ smaller than the maximum supercurrent for zero flux quanta.

The situation is similar when a larger integer number, $N$, of magnetic-flux quanta pierce the junction.  Here, there are $N$ Josephson vortices, each with a bound Majorana state.  For $N\ll\frac{W}{\lambda_B}$, the splitting of these states from tunneling between Majoranas is negligible, except in the vicinity of $\theta_0\approx \pm\pi$ where Majorana states fuse and annihilate at the edges of the junction.  A similar peak-dip current-structure is observed for higher integer number of flux, $N$, with maximal current $I_M\approx \frac{\Delta}{\Phi_0}$ independent of $N$, and with the peak (dip) width $\delta\theta\approx \frac{\pi\lambda_B}{N\ell_B}$.

Like the Majorana zero-modes, as $y_1\rightarrow 0,W$, the non-zero energy Andreev-bound-states on the top and bottom surfaces hybridize and split.  However, unlike the Majorana modes, both branches of the hybridized finite-energy ABS's are occupied and tend to give opposite and cancelling contributions to the Josephson current.

\begin{figure}[ttt]
\begin{center}
\includegraphics[width = 3.2in]{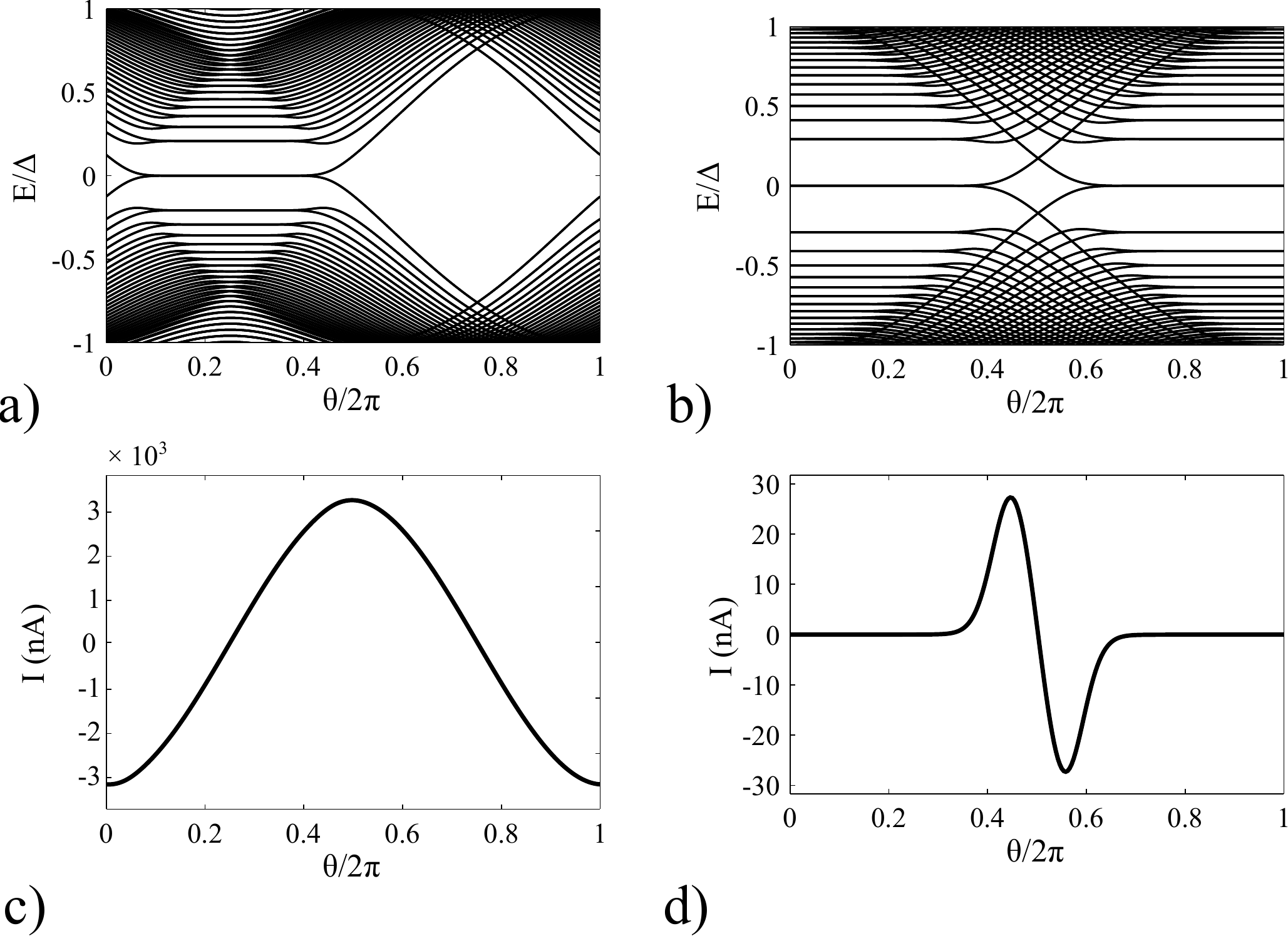}
\end{center}
\vspace{-.2in}
\caption{Numerical computation of the Andreev bound-state spectrum, and Josephson current for Eq.~\ref{eq:heff} with the parameters $v_F = 4.2\times 10^5$m/s, $\mu = 10$meV, $\Delta = 151\mu$eV, and $W = 2\mu$m relevant for Al/Bi$_2$Se$_3$/Al junctions.  Panels a and b show the low-lying Andreev bound-state spectrum as a function of phase difference $\theta$ between the superconductors for $\Phi_B = 0.5\Phi_0$ and $\Phi_B = \Phi_0$ respectively.  The Josephson current corresponding to a) and b) are shown in panels c) and d) respectively.  Panel d) displays the characteristic sharp peak-dip structure from topological Andreev bound-states fusing at the edge of the junction, as discussed in the text.  
}
\vspace{-.2in}
\label{fig:SpectrumandCurrent}
\end{figure}

\vspace{6pt}\noindent{\bf\textit{Numerical Solution - }} To illustrate the predictions of the semiclassical treatment of this simple model, we have also numerically computed the spectrum of Andreev bound-states and their contribution to Josephson current by discretizing Eq.~\ref{eq:heff} on a finite lattice of spacing $a$, with $E(y)\approx \Delta \cos\(\frac{\pi y}{\ell_B}\)$. In this simulations, we used realistic parameters of $v_F$ and $\Delta$ for an Al/Bi$_2$Se$_3$/Al junction, and junction geometry comparable to that in Ref.~\onlinecite{williams,lu, Mason}.  While $\mu$ for Bi$_2$Se$_3$ is typically $\approx 0.1$eV, for simplicity, we illustrate the case for smaller $\mu\approx 10$meV which can readily be achieved by gating\cite{Mason,Sacepe}.  The Andreev bound-state spectrum and corresponding Josephson current are shown in Fig.~\ref{fig:SpectrumandCurrent} for half a magnetic flux quantum (left panel) and a single magnetic flux quantum (right). 

For half a magnetic flux quantum, there is a single Majorana zero-mode when $\theta_0\in\[0,\pi\]$, however, the current is dominated by conventional states.  For a full flux quantum, there are Majorana bound-states for all values of $\theta_0$, which fuse at the junction edge when $\theta_0\approx \pi$ as described above.  Away from $\theta_0\approx \pi$, the spectrum is roughly independent of phase.  Near $\theta_0=\pi$, the ABS's near the $\pi$-crossings on the top and bottom surfaces hybridize and contribute to Josephson current.  However, since the non-zero energy ABS's split in opposite directions the Josephson current is predominately carried by the lowest-energy state, as described above.

The results are similar for larger chemical potential, except that $v_M$ is smaller, leading to a smaller level spacing for the non-zero energy ABS's, and a narrower peak-dip structure at $\Phi_B=\Phi_0$.  Importantly though, the maximum current in the peak-dip structure shown in Fig.~\ref{fig:SpectrumandCurrent}d. is independent of $v_M$, in agreement with the analytic arguments presented above.

\begin{figure}[ttt]
\begin{center}
\includegraphics[width = 2.7in]{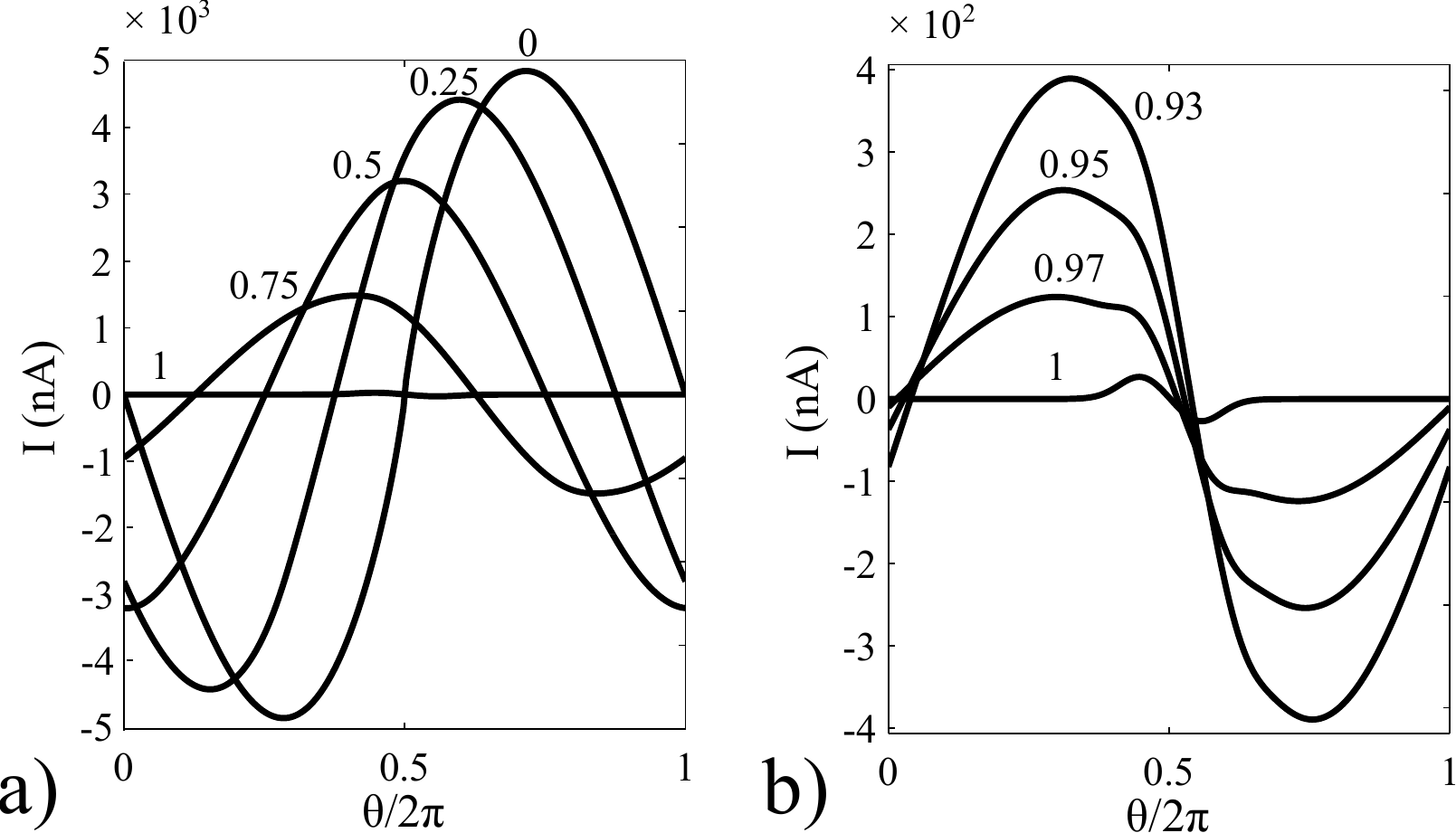}
\end{center}
\vspace{-.2in}
\caption{Current-phase relationship for the parameters listed in Fig.~\ref{fig:SpectrumandCurrent} for a wide range of flux, $\Phi$, (left) and for $\Phi_B$ within a few percent of $\Phi_0$ (right).  Each curve is labeled by value of $\Phi_B/\Phi_0$.  The Majorana contribution to the Josephson current becomes appear as a shoulder in the curves of the right panel which grows and eventually dominates very close to $\Phi=\Phi_0$.}
\label{fig:Current-Phase}
\end{figure}

\begin{figure}[bt]
\begin{center}
\includegraphics[width = 3.4in]{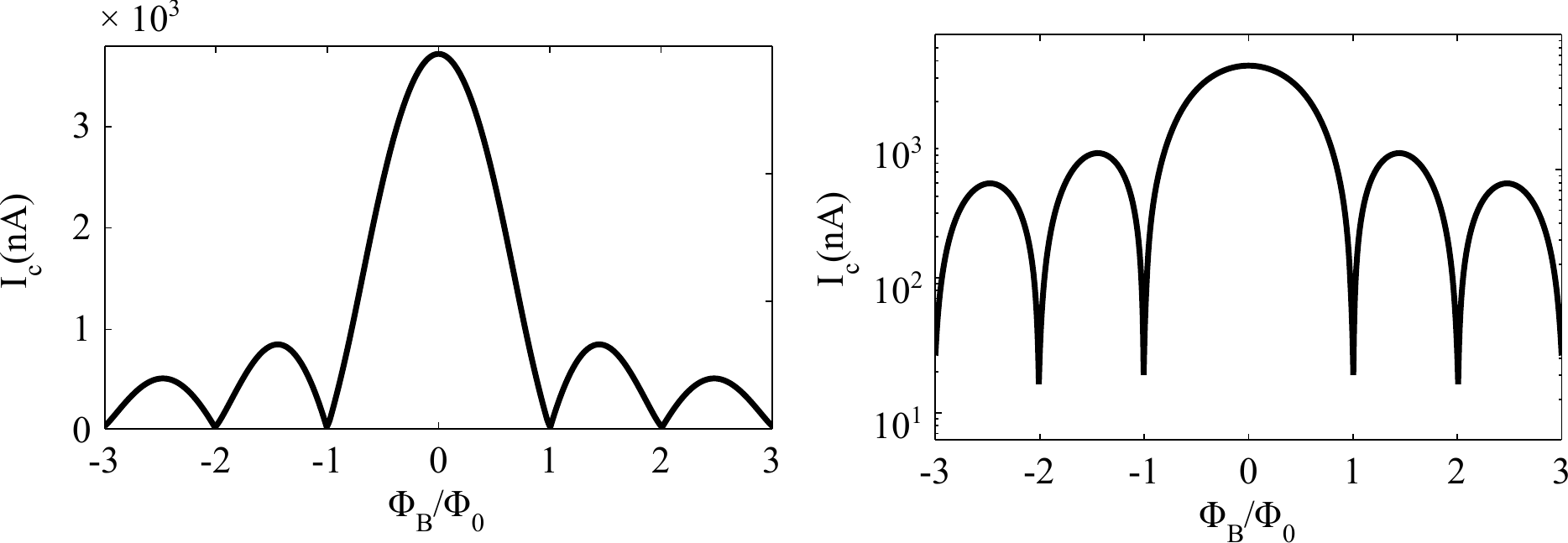}
\end{center}
\vspace{-.2in}
\caption{Critical current, $I_C$, as a function of flux, $\Phi_B$ in linear (left) and log- scale (right).  The curve is quite close to the conventional Frauenhofer pattern, with the exception that $I_C$ does not vanish at integer number of flux quanta due to the extra contribution of the Majorana bound-states (as can be seen in the log-scale plot).}
\vspace{-.2in}
\label{fig:Frauenhofer}
\end{figure}

Fig.~\ref{fig:Current-Phase} shows the current phase-relation for various values of flux through the junction.  The Majorana mode contribution only starts to become visible for flux within $\approx 5\%$ of a single flux quantum (see right panel of Fig.~\ref{fig:Current-Phase}).  This contribution appears initially as a shoulder on the background sine-curve, which strengthens and becomes dominant within $1\%$ of a single flux quantum.

\begin{figure}[tbb]
\begin{center}
\includegraphics[width = 1.7in]{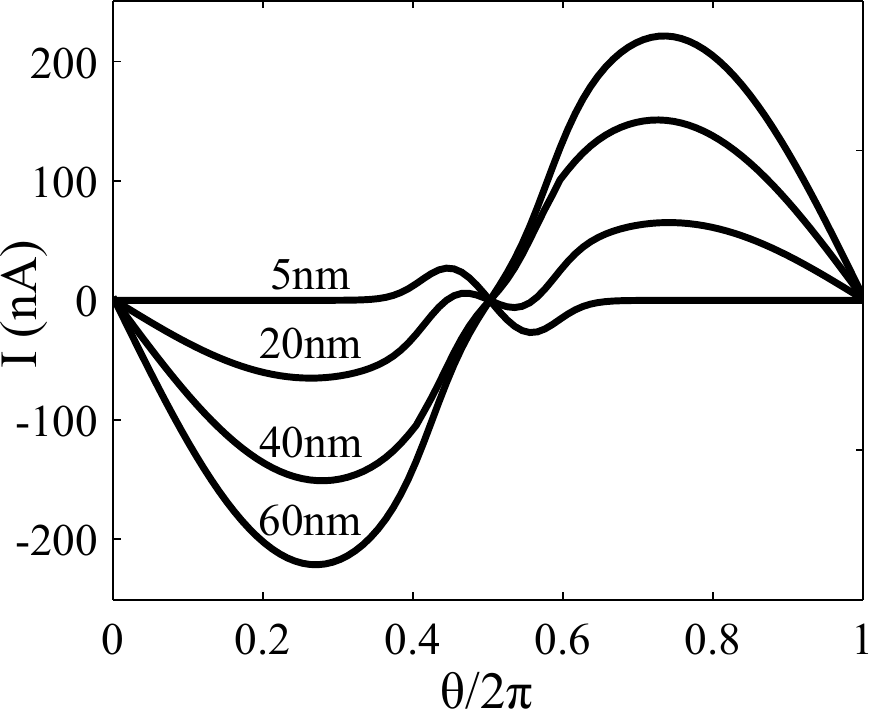}
\end{center}
\vspace{-.2in}
\caption{Current-phase relationship for a single magnetic flux, and varying TI thickness $h=$ 0nm, 20nm, 40nm, and 60nm.  Thicker sides tend to contribute conventional Josephson signatures that mask the topological Andreev-bound-state contributions.}
\vspace{-.2in}
\label{fig:FiniteSides}
\end{figure}

Finally, Fig.~\ref{fig:Frauenhofer} shows the critical current, $I_C$, as a function of flux through the junction.  $I_C$ nearly follows the characteristic Frauenhofer pattern, except that $I_C$ does not vanish for integer flux due to the Majorana contributions to the Josephson current.  In the experiments of Ref.~\onlinecite{williams}, the avoided zero in the Frauenhofer pattern is substantially larger ($\approx 10\%$ of the maximal $I_C$).  As described below (and illustrated in Fig.~\ref{fig:FiniteSides}), this discrepancy can be explained by accounting for the non-negligible thickness of the TI film in these experiments.

\vspace{6pt}\noindent{\bf\textit{Discussion - }}
Up to now we have assumed that the thickness, $h$, of the TI film is smaller than the size of the Majorana states, $\lambda_B$. 
In practice, $\lambda_B$ is at most a few tens of nanometers. 
For thicker films with $h > \lambda_B$, the side of the TI may host additional states compared to $h=0$ 
when the phase difference at the edge is close to $\pi$.  As shown in Fig.~\ref{fig:FiniteSides}, these states will conduct Josephson current, as their energies are sensitive to the phase difference along the sides of the junction.  Like the Majorana contribution described above, the contribution of current from these sides can dominate when there are close to an integer number of magnetic-flux quanta piercing the junction.  Unlike the Majorana contribution however, these states exhibit a more conventional sinusoidal current-phase relationship, rather than a sharp peak-dip structure 
(see Fig.\ref{fig:FiniteSides}).

Therefore, for the purpose of observing Josephson-current signatures of Majorana fermions, it is advantageous to make the TI film as thin as possible without strongly hybridizing the top and bottom surfaces.  Since the confinement length in the z-direction is only a few nanometers, this last constraint is not too severe.  Furthermore, it is advantageous to limit the junction width $W$ (while maintaining $W>\lambda_B$) in order to suppress the contributions from conventional extended states (which scales as $I\sim W$, in contrast to the width independent contribution from the Majorana bound-states).

Lastly, while we have mainly considered the geometry shown in Fig.~\ref{fig:Device}, where superconductors are deposited on the top and bottom surfaces.  In practice it is easier to fabricate devices where superconductors are deposited only on the top surface.  As described above, we expect our analysis to also apply to this simpler to fabricate geometry, so long as the TI thickness is less than a coherence length, and so long as superconductivity is transmitted from top to bottom surfaces via either bulk TI states 
or  boundary states on side surfaces. 
The main difference from the case discussed in the text is that there will be a non-zero hybridization of the ABS's on the top and bottom surfaces through the bulk. 
Nonetheless, in this case the energy splitting of the top and bottom ABS's is position independent, which will not qualitatively alter the Josephson signatures discussed above so long as the splitting is much smaller than the bulk gap $\Delta$.

Throughout this work we have implicitly assumed that the system is in its ground state. Finite temperature will not qualitatively alter our results, provided $T<\delta E_M$.  
Deviations from ground-state behavior may also occur due to sources of single electrons, such as localized impurity states near the junction.  When the Majorana states are far from the junction edge, their mutual fermion parity can be switched by tunneling to local single-electron sources.  After such a parity switching, the Majorana modes will follow the positive energy branch as they approach the junction boundary and fuse, thereby contributing the opposite sign of Josephson current compared to the equilibrium case discussed above.  Such parity switching events can thereby lead to hysteretic current-phase behavior, whose observation would provide strong evidence for the Majorana character of the Andreev bound-states in the junction.

\noindent\textit{Acknowledgements - } We thank Pouyan Ghaemi, Patrick A. Lee, Joel Moore and especially Bertrand I. Halperin for useful discussions. ACP acknowledges funding from DOE Grant No. DEFG0203ER46076, and also thanks the Kavli Institute for Theoretical Physics at UC Santa Barbara, supported in part by the NSF Grant No. NSF PHY11-25915, for hosting part of this research.


\begin{thebibliography}{10}
\bibitem{Footnote:FluxConcentration}
Flux expulsion from the superconductors can concentrate flux into the junction region.  Consequently the magnetic flux through the junction, $\Phi_B$, defines an effective area: $A_\text{eff} = \Phi_B/B$ which is in general different than the geometry area $W\times L$.
\end{thebibliography}
\end{document}